\documentstyle[manuscript,aps,version2]{revtex}
\begin{document}
\draft

\title
\bf  

A Simple Model of Large Scale Organization in Evolution

\endtitle

\author{S. C. Manrubia $^1$ and M. Paczuski$^{2,3}$}
\instit
$^1$CSRG, Departament de F\'isica i Enginyeria Nuclear, UPC \\ 
c/ Sor Eul\`alia d'Anzizu s/n 
 Campus Nord, B4-B5. 08034 Barcelona, Spain \\
$^2$Department of Physics, Brookhaven National Laboratory, Upton NY 11973 \\
$^3$Department of Physics, University of Houston, Houston TX  77204-55063\\

\endinstit

\medskip

\abstract 

A mathematical model of interacting species filling ecological niches
left by the extinction of others is introduced.  Species organize
themselves into genera of all sizes.  The size of a genus on average
grows linearly with its age, confirming a general relation between Age
and Area proposed by Willis.  The ecology exhibits punctuated
equilibrium.  Analytic and numerical results show that the probability
distribution of genera sizes, genera lifetimes, and extinction event
sizes are the same power law $P(x) \sim 1/x^2$, consistent with
paleontological data.

\endabstract


Many years ago, Willis noted that genera could be composed of
many species or of only one; he noted that similar regularities
in the statistical properties of genera
occured whether one is studying flowering plants or e.g. beetles.
Attempting to formalize these observed regularites, he
 postulated a relation between the age and size
of a genus, ``Age and Area,'' which states that older genera on
average include more species than younger ones \cite{willis}.  With
Yule he noted a power law relation for the number of genera with $s$
species, $P_{gen}(s) \sim s^{-\tau}$ with $\tau$ approximately 2
\cite{willis,yule}.  Recently, Burlando \cite{burlando} observed
scaling behavior across the taxonomic hierarchy
 also giving $\tau \simeq 2$ (see in addition
Ref. \cite{adami}).  Similarly, the distribution of life times, $t$,
of fossil genera \cite{raup} can be described by a power law
$P_{life}(t) \sim 1/t^{\tau_t}$ with $\tau_t \simeq 2$ \cite{sfbj}.
When viewed at sufficiently large time scales, the pace of extinction
itself is episodic with long periods of stasis interrupted by sudden
bursts of mass extinction \cite{gould}.  Punctuated equilibrium with
scale-free extinctions has been attributed \cite{kauffman} to the
self-organized critical \cite{soc} dynamics of strongly interacting
species, without the need for catastrophic exogenous causes such as
meteorites.  The punctuated equilibrium process governing large scale
evolution may be sufficiently robust or universal to be captured by an
abstract, mathematical model.  Bak and Sneppen \cite{BS} introduced a
self-organized critical model of coevolving ``species'' where the
least fit undergoes pseudo-extinction and affects the fitness of other
species in the ecology, leading to extinction events of all sizes.
However, their dimension-independent lifetime distribution
$P_{life}(t)\sim 1/t$ \cite{sfbj,commenta} is in disagreement with Raup's
\cite{raup,sfbj} paleontological data.  More importantly, their model
lacks any emergent taxonomic structure, which is an essential part
of large scale organization in evolution.

Here we show that these three, seemingly unrelated distributions for
extinction event sizes, genera sizes, and genera lifetimes, which
characterize the large scale behavior of evolution, can be unified in
terms of a simple mathematical process.  We introduce an abstract
model for large scale evolution and study it both analytically and
numerically.  In our model, surviving species can diversify into
ecological niches left by previous extinction.  This is
implemented in terms of a Polya urn type of process.  All species are
subject to a general drift over time to lower viability which
eventually leads to their extinction.  This is consistent with the
view that most changes in the ecology have a deleterious effect on
currently existing species.   In addition, due to
interactions between species they experience changes in
their viability, either favorable
or unfavorable, arising from previous extinction of other species in
the ecology.  Species in our model organize themselves into genera of
all sizes.  The emergent genera obey a general Age and Area
\cite{willis} relation which we find is linear ($s\sim t$) in contrast
to Yule's \cite{yule} conjecture $ \ln s \sim t$.  This large scale
organization of species provides a simple mechanism which shapes all
three probability distributions into a power law $P(x) \sim 1/x^2$,
where $x$ is the extinction event size, genera lifetime, or genera
size.  In all three cases the $1/x^2$ behavior is consistent with
previously reported paleontological data
\cite{raup,sfbj,kauffman,burlando,willis}, indicating that our model
may plausibly describe a universality class sufficiently broad to
include real evolution. It can be tested further
by directly comparing the age and size of extinct genera with our
result that on average $s \sim t$.

Our model was inspired, in part, by considering a more complicated
``connection'' model introduced by Sol\'e and Manrubia \cite{sole}.
The advantage of our model is that it is extremely simple and robust.
Its simplicity makes it easier to
study large systems numerically;  it is also analytically tractable.
That such a simple model exists which describes a process giving
large scale organization in evolution together with the above mentioned
distributions is significant we think because it illustrates 
a potentially universal mechanism that would apply even beyond
the context of biological
evolution discussed here.

We begin by briefly describing the
connection model.  An $N\times N$ interaction matrix $W$ defines the
interaction, either favorable or unfavorable, between $N$ objects that
represent species.  For a species $i$, the output elements $W_{ij}$
define its affect on the other species $j$, while its viability is the
sum of its input elements $v(i)=\sum_j W_{ji}$.  If $W_{ji}>W_{ki}$
then species $j$ has a more beneficial effect (and species $k$ has a
more deleterious effect) on the ability of species $i$ to survive.  If
the viability $v(i)<0$, then species $i$ goes extinct, and
the connection
elements of the rows and columns for that extinct species are replaced with
a copy of the corresponding elements of another surviving species.
This copying in turn changes the viability of other species,
and leads to a chain reaction of extinction events.  The system is
driven by slow random changes in the matrix $W$ which tend
to lower the viability of the surviving species, and slowly
differentiate copies from each other leading to speciation.  They
observed that the connection model exhibits extinction events of all
sizes where the species that go extinct together tend to be recent
copies.

In our one-dimensional model we assign to each of $N$ particles that
represent species an integer viability $v(i)$.  The dynamics consists
of three steps as illustrated in Fig. 1: (1) species drift
stochastically to lower viability; (2) species with viability below a
threshold $v_c$ become extinct.  The extinct species are each replaced
with a ``daughter'' speciation of a surviving species.
This is the Polya urn mechanism in our model; (3) Due to interactions
between species the surviving
species receive a change in their viability resulting from the
extinction event.  Specifically, at each time step the following
operations are performed in parallel for all species $(i)$: (1) with
probability $1/2$, $v(i)=v(i)-1$; otherwise $v(i)$ is unchanged; (2)
for each $i$ such that $v(i)<v_c$ a surviving species $(j)$ with
$v(j)\geq v_c$ is selected at random and $v(i)=v(j)$.  This step
represents a speciation event where one species branches into two. 
(3) all $N-s$ species that survived extinction receive a
coherent influence $q(s)$, so that $v(j)=v(j) + q(s)$.  After an
extinction event of size $s$, $q(s)$ is chosen from the uniform
distribution $-s \leq q(s) \leq s$.  Thus, only large extinctions can
cause large subsequent changes in the ecology.  The form of 
$q(s)$ is elaborated on later.  It is important to note that, unlike
the connection model, our model's behavior is robust with respect
to varying the parameter $v_c$, since the entire system is translationally
invariant in viability.

Our model can be viewed as an example of transport in one dimension,
where particles are conserved.  In the steady state the smooth drift
of species toward lower viability will be balanced by the intermittent
replacement of extinct species with speciations of surviving ones,
which by definition have higher viability.  The average viability in
the system ${\bar v}= {1\over N}\sum_i v(i)$ exhibits stick-slip
behavior as shown in Fig. 2, similar to the behavior observed in the
connection model \cite{sole}.  

Due to replacement of extinct species with speciations of surviving
 ones, the species tend to form groups with similar viability, which
 drift and diffuse together toward the extinction threshold.  The state of the
 system may be characterized by $n(v,t)$, the number of species of
 viability $v$ at time $t$, where $\sum_v n(v,t)=N$.  A snapshot of
 the system is shown in Fig. 3.  At a microscopic scale, $n(v,t)$ is
 peaked with well defined bumps that give rise to the
 temporally intermittent sequence of extinctions as shown in the
 insert of Fig. 2.

 We can identify all species within each bump as members of the same
genus for the following reasons: Each viability bump is separated from
the others by an empty interval where $n(v,t)=0$.  Since these empty
intervals cannot be filled by the replacement of extinct species with
speciations of surviving ones, the dynamics tends to maintain the
sharp separation between different bumps.  Therefore, they are
long-lived metastable entities.  By making a histogram of the genera
sizes, or area under each bump, observed in snapshots at spaced time
intervals we find a power law for the number of species within each
genus as shown in the insert of Fig. 3 with an exponent $\tau \simeq
2$.  Also, the total number of genera in the system displays an
intermittent pattern of diversification (increase) and contraction in
time, qualitatively similar to real data \cite{realdata}.  

The age of a newly created species following extinction is set to
zero, and incremented by one unit at each step in the simulation.  The
age of a genus is the age of the oldest species in the corresponding
bump.  When a bump passes through the extinction threshold $v_c$, we
measure its age $t$ and size $s$ (or area).  The distribution of sizes
of extinct genera is the same, within numerical accuracy, as the
snapshot distribution described above.  We numerically determined the
relation between the age and size of extinct genera in a system of
size $N=1000$ including $10^7$ time steps, and found a linear relation
$t=ms$, with $m\simeq 0.6$.  This numerical result indicates
that emergent genera on average grow at
constant rate irrespective of their size.  The numerical result 
$\tau=2$, then implies $\tau_t=2$, in agreement with real data.  Data
collapse of the distribution of extinction event sizes for different
system sizes also indicate a power law with exponent $\tau_{ext}=2$,
as shown in Fig. 5.  Note that at the beginning of the numerical
simulation, with random initial conditions, there are only small
extinctions and small genera.  Thus the power law distributions
observed in the steady state are ``emergent''; they are
consequences of the self-organized critical dynamics of our model.

The first step in our model, drift to lower viability, takes into
 account slow random mutations of the matrix elements $W_{ij}$ in the
 connection model that tend to lower the viability of all species.
 This slow external driving, similar to that used in earthquake models
 \cite{christ}, represents the cumulative effect on species of small
 changes in the environment, which we propose tend to make species
less able to maintain their population over time.
  The second step represents true
 extinctions of species.  The third step represents the effect of these
 extinctions on the surviving species.  For more details see
 Ref. \cite{comment1}.  Our preliminary results indicate that the
 connection model also exhibits emergent genera with a broad size
 distribution, intermittent diversification, as well as an Age and
 Area relation.

We now  discuss the analytic results for the
transport model.  In the stationary state, $P_{ext}(s)$ is the probability
distribution to have an extinction event (avalanche) of size $s$ and
$G(q)$ is the probability distribution to have an influence
 of size $q$.
These distributions are {\it self-consistently} related via
\begin{eqnarray}
G(q)&=& \sum_{{\rm all \ } s \geq q} {P_{ext}(s) \over 2s +1} 
\label{e1} \\
P_{ext}(s)&=& \sum_{{\rm all \ } q} G(q) 
\delta\Big(\sum_{v=0}^{q-1}{\bar n}(v),s\Big) \quad ,
\label{e2}
\end{eqnarray}
where $\bar n(v)$ is the time averaged viability profile in the steady
state.  The first equation is exact.  The second assumes that the
avalanche distribution comes from influences, $q$, on  the time average
viability profile, rather than the actual time dependent profile.  The
extinctions and influences are treated in terms of their full
probability distributions, while the viability profile of species is treated
only in terms of its average.  This can be justified {\it a
posteriori} in terms of a separation of scales argument similar to
singular diffusion \cite{singular}.  

Next, we assume that the cumulant of $\bar n(v)$ is not singular
around $v=0$, so that it has a Taylor series expansion.  In 
the interval $1 \ll q \ll N$, where  $N
\rightarrow \infty$, $\int^q \bar
n(v) dv= Aq + ...$.  Combining Eqs. (1,2) with the Taylor series expansion
gives
\begin{eqnarray}
{dG(q) \over dq}\bigg|_q = {-1 \over 2Aq} G\Bigl({q\over A}\Bigr) \qquad 
{\rm for} \qquad 1 \ll q \ll N \quad .
\end{eqnarray}
It is easy to show that Eq. (3) has a scaling solution $G(q) \sim
q^{-\tau_{ext}}$ where $\ln (2\tau_{ext}) = ({\tau_{ext}-1})\ln A$.  Also
the avalanche and influence distributions are asymptotically the same;
$P_{ext}(q)\sim G(q)$.

The steady state equation for the time average profile is
\begin{equation}
{\bar n}(v)= {1\over 2} \sum_{s=0}^{N-1} \sum_{q=-s}^s {P_{ext}(s)\over 2s+1}
\Big({\bar n}(v+q)+ {\bar n}(v+q+1)\Big)  +
{1\over 2}\Big({\bar n}(v) + {\bar n}(v+1)\Big)
\sum_{s=0}^{N-1} {s P_{ext}(s) \over N-s} \quad .
\end{equation}
For large $N$, we try the solution $n(v)=n_o e^{-c\, v/N}$ and find to
leading order in $N$
\begin{equation}
1= {1\over 2}\sum_{s=0}^{N-1}\sum_{q=-s}^s {P_{ext}(s)\over 2s+1}e^{-cq/N}(2-c/N)
+\sum_{s=0}^{N-1} {s P_{ext}(s) \over N-s} \quad .
\label{e5}
\end{equation}
Expanding for $q<<N$, the $q=0$ part of the first term on the right
hand side gives $1- {c\over 2N}$.  The leading part cancels the number
one on the left hand side of Eq. \ref{e5}, and the negative remainder
which comes from the drift must be cancelled by the remaining terms in
the equation.  Completing the expansion in $q$, only the even terms
survive the symmetric sum over $q$. These terms are all positive as is
the last term in Eq. \ref{e5}.  From Eq. (3) 
all of these positive terms scale
$\sim N^{1-\tau_{ext}}$.  Only when $\tau_{ext}=2$ can the positive
terms cancel the only negative term $({-c\over 2N})$.  In this case,
a consistent solution exists for the exponential profile.  Our
numerical simulation results show that the average profile is
indeed exponential with $c \simeq 6$ with $A \simeq 4$, both confirming
$\tau_{ext}=2$.

Note that our theory thus far has removed genera bumps by only
treating the time average profile.  The weak $\ln N$ divergence of both
the average size of influences and average size of extinction events
justifies our separation of scales assumption for large $N$.  Since
the shifts are small relative to $N$, only a finite number of genera
on average pass the extinction threshold following an extinction.  Then
$\tau=\tau_{ext}=2$, in agreement with the numerical simulation
result.  Finally, previous interpretations of available data from the
fossil record for extinction size distributions, genera (and higher
order taxa) abundance distributions, and lifetime distribution of
genera are consistent with our unified result in terms of a simple
model that they are each decaying power laws with exponent 2.

This work was supported in part by the
U. S. Department of Energy under Contract No. DE-AC02-76-CH00016 and
DE-FE02-95ER40923 and by a grant of the Spanish Government DGYCIT 1995 
PB94-1195.  We thank P. Bak, M. Goldhaber, and
 R. V. Sol\'e for interesting discussions about evolution.
SCM  acknowledges the hospitality 
of Brookhaven National Laboratory, where part of this work was done.

\figure{\label{picture} Dynamics of the model.  The horizontal axis is
the viability and the blocks represent species.  The dotted line is
the extinction threshold.  (0) Initial configuration.  (1)
Leftward stochastic drift. (2) Extinction and replacement.
(3) Coherent influence to the survivors.  Here the extinction had size
$s=5$ and the influence of the extinction
had value $q=-2$.  The species that move
at each step are shaded.  }

\figure{\label{viability} The average viability as a function of time
in a system of size $N=1000$ exhibiting stick-slip dynamics.  The
steep jumps, or slip events, are followed by slow relaxation to the
threshold for extinction.  The insert shows the temporal sequence of
extinction event sizes over the same interval.}

\figure{\label{bump} A snapshot of the viability profile $n(v,t)$ in
a system of size $N=400$.  The pattern is intermittent with both small
and large bumps.  The insert
shows the distribution of genera sizes averaged over a total time
interval of $10^7$ steps with a snapshot taken every 100 time steps
for a system of size $N=1000$.  The curve can be described as
 power law with a cutoff
at the system size.}

\figure{\label{datacollapse} Data collapse result.
$P(s,N)$ is the probability
to have an extinction event of size $s$ in a system of size $N$.  The
plateaus for different system sizes show that $P(s,N)= F(s/N)/s^2$
where $F(x)$ is simple scaling function which is constant for $x \ll
1$ and approaches zero as $x \rightarrow 1$.  This agrees with our analytic
result that $\tau_{ext}=2$.}


\begin{references}
\bibitem{commenta} The lifetime distribution in units of simulation time
is different than the physically relevant time, which is argued to
be exponential in fitness.  This gives the universal behavior $1/t$,
 even though
the behavior in simulation time does depend on dimension.
\bibitem{comment1} Because they are close copies, the species which go
extinct together in the connection model have similar output
connections to any other species.  Thus, the change in viability of
any surviving species $j$ due to $s$ species going extinct involves
the removal of $s$ correlated numbers contributing to the viability
$v(j)$.  This is why we use an influence distribution bounded by $s$,
rather than e.g. $\sqrt s$.  Using the $\sqrt s$ distribution gives
$\tau_{ext}=1$.  However, breaking the plus/minus symmetry of the
influence distribution has no effect on the critical behavior as long
as some influences are favorable, increasing viability.  Note also
that the value of $v_c$ in our model completely arbitrary, unlike the
case for the connection model which is not critical unless
$v_c=0$.   Here we use $v_c=0$.
In our model we use the same influence following extinction for all
surviving species, rather than a different one for each species.  In
the connection model, not only are extinct species closely related,
but we observe that the surviving species also form groups with
similar matrix elements.  Within each group or genus, all species will
have similar connection to any other species, so that all species in
the same genus get virtually the same effect in their
viability from an extinction of
another genus.  In our model, we approximate the behavior of the
connection model by assuming there is only one surviving genus group
rather than many.
\end{references}
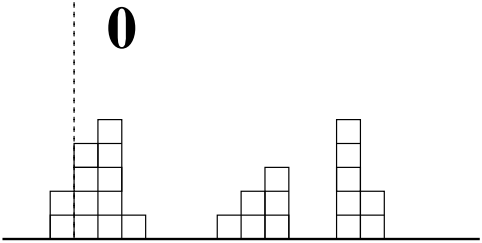
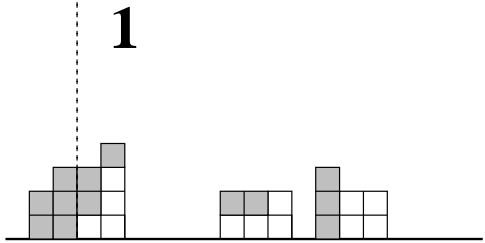
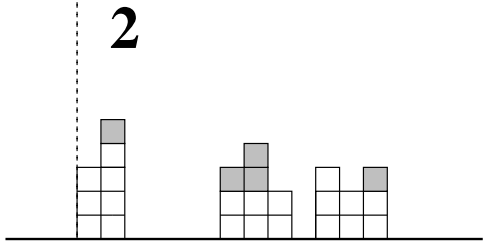
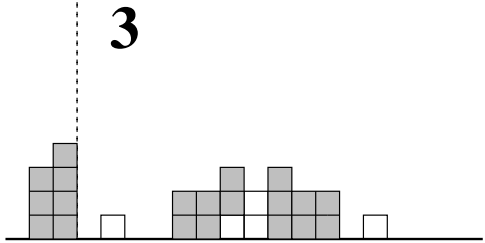
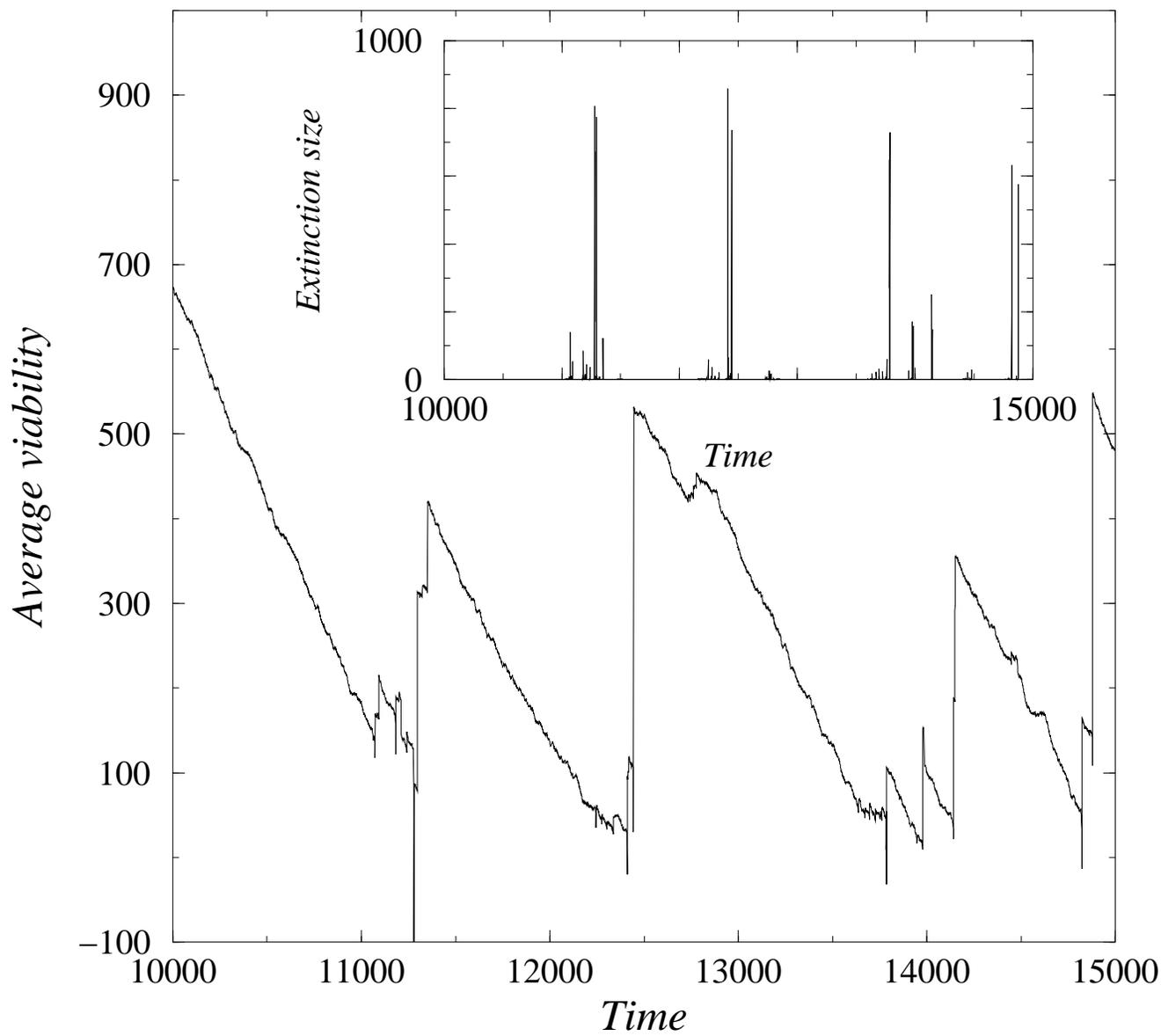
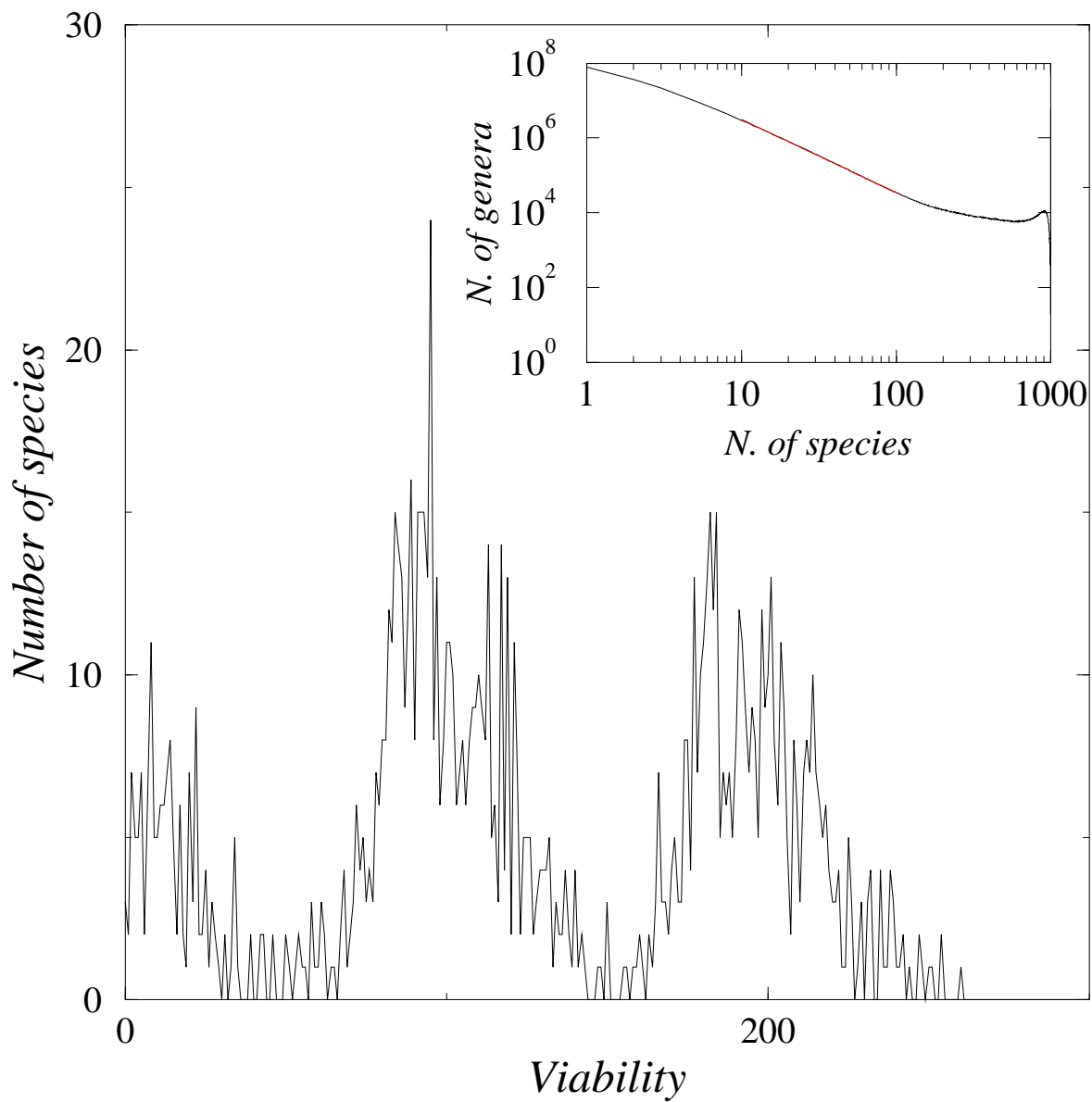
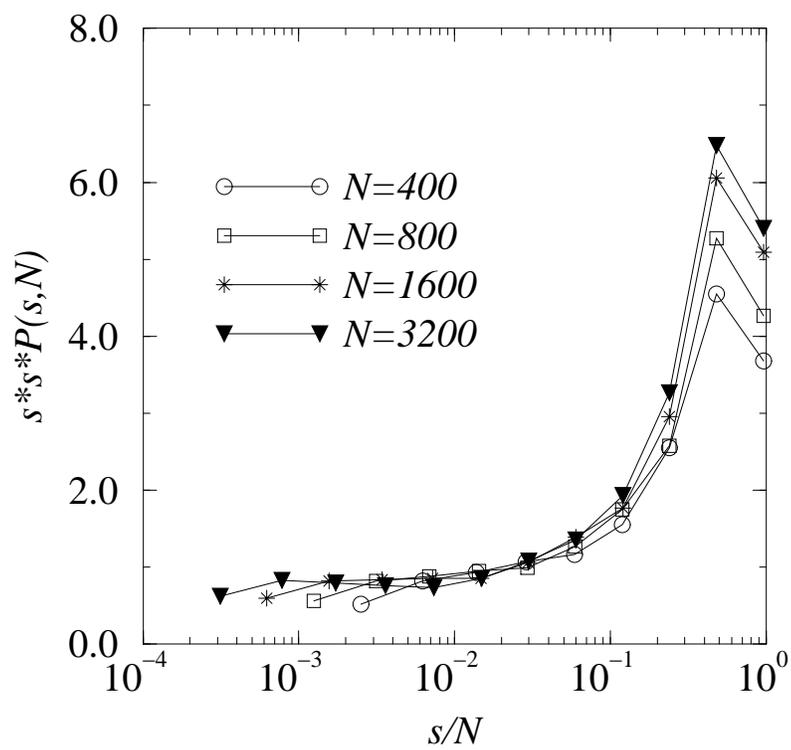
\end{document}